# Upper Critical Field of a Novel 18-K Superconductive Phase in Metastable Yttrium Sesquicarbide Prepared Using High Pressure Technique

T. Nakane,[1] T. Mochiku,[1] H. Kito,[2] M. Nagao,[1] J. Itoh,[2] H. Kumakura,[1] and Y. Takano[1]

1. *National Institute for Materials Science, 1-2-1 Sengen, Tsukuba, Ibaraki 305-0047, Japan*
2. *National Institute of Advanced Industrial Science and Technology, 1-1-1 Umezono, Tsukuba, Ibaraki 305-8568, Japan*

## Abstract

The metastable phase of an yttrium sesquicarbide compound, which was vary recently reported as a new superconductive phase with a $T_c$ value of 18 K, was prepared under high pressure from the nominal composition of $Y_2C_3$ and $Y_2C_{2.9}B_{0.1}$ in order to investigate this phase. We have succeeded in reproduction, and in the transport measurement for this 18-K phase. It is considered that the reason for the higher $T_c$ value than the 11.5 K reported previously is not the small amount of B as the impurity element from the BN crucible. The $B_{c2}$ value at 0 K estimated from the resistivity data for the 18-K phase seems to be over 30 T. It means that the $T_c$ and $B_{c2}$ values of the 18-K phase are as high as those of the general A15 compounds.

.

The discovery of new superconductive materials with ever-higher transition temperatures, $T_c$, has great potential and is accelerating the practical application of superconductors. Very recently, Amano *et al*. reported the discovery of a novel superconductive phase with a $T_c$ value as high as 18 K in an yttrium carbide compound (18-K phase) prepared by a high-pressure synthesis technique.[1] The chemical composition and crystal structure of this phase are still not well understood because of the instability of the sample in air and the difficulty synthesizing a single-phase sample. Amano *et al*. assumed that this superconductive new phase originated in the $Y_2C_3$ phase with a cubic $Pu_2C_3$-type crystal structure. In previous work, the $T_c$ value of this yttrium sesquicarbide system was reported to be 11.5 K (11-K phase).[2] However, it is also known that the superconductive property of this yttrium sesquicarbide system is sensitive to the carbon content.[2,3] Moreover, the highest $T_c$ value of this structure was ≈ 17 K reported for the $(Y_{0.7}Th_{0.3})C_{1.55}$ phase.[3] Furthermore, there have also been many reports regarding the discovery of high-$T_c$ superconductive phases in boride and/or carbide in the past few years.[4-8] Hence, the report for the 18-K phase is believed to be reliable. A slight difference in the experimental technique may optimize the $T_c$ value of this system without any element of impurity. However, the effect of the partial substitution of B for the C site is also considered as the origin of the 18-K phase, since Amano *et al.*[1] used a BN crucible for high-pressure synthesis and Krupka *et al.* did not.[2] Therefore, demonstrating the reproducibility by the other group seems to be very important for the new superconductive 18-K phase in an yttrium sesquicarbide system. The superconductive upper critical field, $B_{c2}$, is also of high interest from the viewpoint of practical applications even if the 18-K phase is synthesized under high pressure. Furthermore, an investigation of these materials will ultimately provide guidelines for designing superconductive materials.

Now, the synthesis method to obtain a single-phase sample of this 18-K phase has not been yet established, yet. Because this phase is unstable in air, some physical investigations are difficult. One such investigation concerns the transport measurement.[1] However, the transport measurement seems possible when the process of contacting the wire is made in non-air atmosphere. The transport measurement makes it possible to evaluate not only the $T_c$ value, but also the $B_{c2}(T)$ line for a multi-phase sample. Therefore, the superconductive new 18-K phase in the yttrium sesquicarbide system is being investigated to demonstrate the reproducibility and to evaluate the $B_{c2}(T)$ line by transport measurement.

Two starting materials, yttrium metal (99.9%) and carbon (graphite: 99.9%) powders, were mixed with a slightly carbon-rich ratio of Y : C = 2 : 3.01. In addition, the B-substituted starting material was prepared with a nominal composition of $Y_2C_{2.9}B_{0.1}$ using the B powder (99.9%). These mixed powders were reacted at 1400 $^{o}$C for 30 minutes under 3.5 GPa using a cubic-anvil-type high-pressure apparatus in the BN crucible. The as-synthesized sample was characterized for the magnetization by SQUID magnetometer, for electrical resistivity by a four-probe resistive method and for the phase analysis by an X-ray diffraction (XRD) pattern. Because of the instability of the as-synthesized phase in air, the SQUID measurement was performed without removing the sample from the BN crucible, and contact with the Au wire for the transport measurement was made with the sample in a dry He atmosphere.[9] The XRD data were obtained from samples that had been polished with sandpaper and coated with grease.[10] The $B_{c2}$ values for the 18-K phase and for the 11-K phase are estimated from the on-set $T_c$ values of resistivity data in an applied magnetic field. Moreover, the values of the irreversibility temperature, $T_{irr}$, and irreversibility field, $B_{irr}$, were evaluated from the magnetization data as the cross point of the zero-field-cooling and field-cooling curves and the hysteresis curve, respectively.

The high-pressure synthesis technique was used to obtain the superconductive 18-K phase in the yttrium sesquicarbide compound with a nominal composition of $Y_2C_3$. Although the sample took on a dark-gold color after being polished with sandpaper, it changed to a light gray powder within several seconds on the surface in air. The resistivity drastically increased around 250 K for this sample. The absolute value of resistivity at 300 K, *R*(300), is 3 times as high as the *R*(250), even if the *R*(30) is 0.8 times as high as the *R*(250). The reason for the deterioration is speculated to be due to the reaction with water in air.

Figure 1 shows the XRD pattern of the sample before deterioration. Almost all peaks are assigned as reflections from the $Y_2C_3$ phase with the $Pu_2C_3$-type structure, but peaks due to the Y metal are also found. This result suggests that the $Y_2C_3$ phase is decomposed into the Y metal and acetylene gas after the reaction with the water in air. An effort was made to measure the XRD pattern of the deteriorated powder. However, distinctive peaks in the powder could not be obtained.

The temperature dependences of the magnetization and the electrical resistivity are shown in Figures 2 and 3, respectively. The on-set $T_c$ values of the 18-K phase obtained from the magnetization and the transport measurements are 17.3 K (Figure 2) and 17.9 K (Figure 3), respectively. These values are consistent with the data reported by Amano *et al*.[1] The superconducting transition is not sharp for these data. It seems to be due to the existence of some impurity phases and deteriorated grain boundaries owing to the instability of the sample. The temperature dependence of the magnetization data for the B-substituted sample with the nominal composition of $Y_2C_{2.9}B_{0.1}$ is also included in Figure 2. The $T_c$ value of the $Y_2C_{2.9}B_{0.1}$ sample ($\approx$ 16.4 K) is lower than that of the non-substituted one. It shows that the improvement of the $T_c$ in $Y_2C_3$ phase does not originate in the partial substitution of B for the C site, or that the optimum composition of B for obtaining a $T_c$ value as high as 18 K is at least less than 0.1. Therefore, we focused on the non-substituted yttrium sesquicarbide compound for the investigation of the 18-K phase.

Figure 3 shows the two-step superconducting transition that occurs as a result of the coexistence of the 18-K phase and the other phase (11-K phase). We also found a trace of a low $T_c$ phase with the $T_c \approx 4$ K in the resistivity measurement without an applied magnetic field. It originated in a small amount of $YC_2$ [4] as an impurity phase. Contamination of the 11-K phase, which was reported by Krupka *et al* in Ref. 2, was observed in the resistivity data although it could not be identified by the magnetization measurement. This means that the 11-K phase exists within the 18-K phase, which shields the 11-K phase from the applied magnetic field during the magnetization measurement. On the other hand, the zero-resistivity could not be measured for the sample used in this study, and the residual resistivity of the data measured in 0 T seems to be quite high. Therefore, it is assumed that the volume fraction of the 18-K phase is small.

The $B_{c2}$ values for the 18-K and 11-K phases are plotted in Figure 4. The $B_{c2}(T)$ line shows a positive curvature at a low magnetic field, which is similar to that in borocarbide systems[11] and $MgB_2$.[12] Except for this region, the $B_{c2}(T)$ line of the 18-K phase is linear with a gradient of $-dB_{c2}/dT = 2.52$ T/K. This gradient is much larger than that for $MgB_2$ ($\approx 0.74$ T/K).[12] On the other hand, the gradient of the 11-K phase is 1.80 T/K, which is smaller than that of the 18-K phase. By extrapolating these lines, we estimate the $B_{c2}$ values at 0 K for the 18-K phase and the 11-K phase to be 43.5 T and 18.3 T, respectively. On the other hand, the $B_{c2}(0)$ is given by the formula, $B_{c2}(0) = -0.6913(dB_{c2}/dT)T_c$, for the type-II superconductor in the dirty limit.[13,14] According to this equation, the $B_{c2}(0)$ values of the 18-K and 11-K phases are 31 T and 14 T, respectively. These experimental data show that the $B_{c2}(0)$ value of the 18-K phase is as high as those of the A15-type intermetallic compounds.

The values of the $T_{irr}$ and the $B_{irr}$ are plotted in Figure 4. The irreversibility lines are consistent. When the irreversibility line is extrapolated linearly, the $B_{irr}(0)$ value of $\approx 18$ T was obtained. The irreversibility line of this compound seems to lie on quite a low field. However, the volume fraction of the 18-K phase is considered to be small according to the data shown in Figure 3. Hence, it is assumed that this property is improved by purifying the sample.

In summary, we have synthesized the 18-K phase in an yttrium sesquicarbide compound, which Amano *et al.* reported as a new superconductive phase,[1] by utilizing a high-pressure technique. It is a metastable phase in air owing to the reaction with water. A B-substituted sample with a nominal composition of $Y_2C_{2.9}B_{0.1}$ was also prepared in order to check the possibility of the partial substitution of B for C as contamination from the BN crucible. However, the $T_c$ value decreased by this substitution. Therefore, the reason for the formation of the 18-K phase is not due to the small amount of B as the doping element, or the optimum composition of B for obtaining the 18-K phase is less than 0.1.

The resistivity data for the superconductive phase were successfully obtained. The $B_{c2}(0)$

value estimated from the resistivity data for the 18-K phase seems to be higher than 30 T. The two magnetic irreversibility lines estimated from the $T_{irr}(B)$ and $B_{irr}(T)$ are consistent; however, they are in a much lower magnetic field than the $B_{c2}(0)$ line. The resistivity data implies that the phase connectivity between the superconducting grains is weak and the fraction of the 18-K phase is small. Hence, it is assumed that the $B_{irr}(T)$ line is possibly improved by purifying the sample.

According to the experimental data, the $T_c$ and $B_{c2}$ values of the 18-K phase are as high as those of the general A15 compounds. The 18-K phase is a very rare compound with high potential for applications, except for A15 compounds. Therefore, the investigation for the 18-K phase is expected to provide valuable information for obtaining a new superconductor with the potential for applications. The phase identification of the 18-K phase and the establishment of the synthesis technique for a single-phase sample and a single-crystal sample are desired. Moreover, the establishment of the synthesis technique preparing the 18-K phase under a normal pressure atmosphere is also strongly desired from the viewpoint of the applications.

The authors would like to thank Prof. J. Akimitsu of Aoyama-Gakuin University for supplying valuable information. We also acknowledge Dr. H. Fujii, Dr. H. Takeya, and other members of the National Institute for Materials Science for their kind support and helpful suggestions.

**Figure Captions**

**Fig. 1** X-ray diffraction pattern of the as-synthesized sample.

**Fig. 2** Temperature dependence of the magnetization for the as-synthesized sample of the yttrium sesquicarbide phase.

**Fig. 3** Temperature dependence of the electrical resistivity in the applied magnetic fields for the as-synthesized sample of the yttrium sesquicarbide phase.

**Fig. 4** Magnetic phase diagram obtained from the magnetization and transport measurement for the as-synthesized sample of the estimated yttrium sesquicarbide phase.

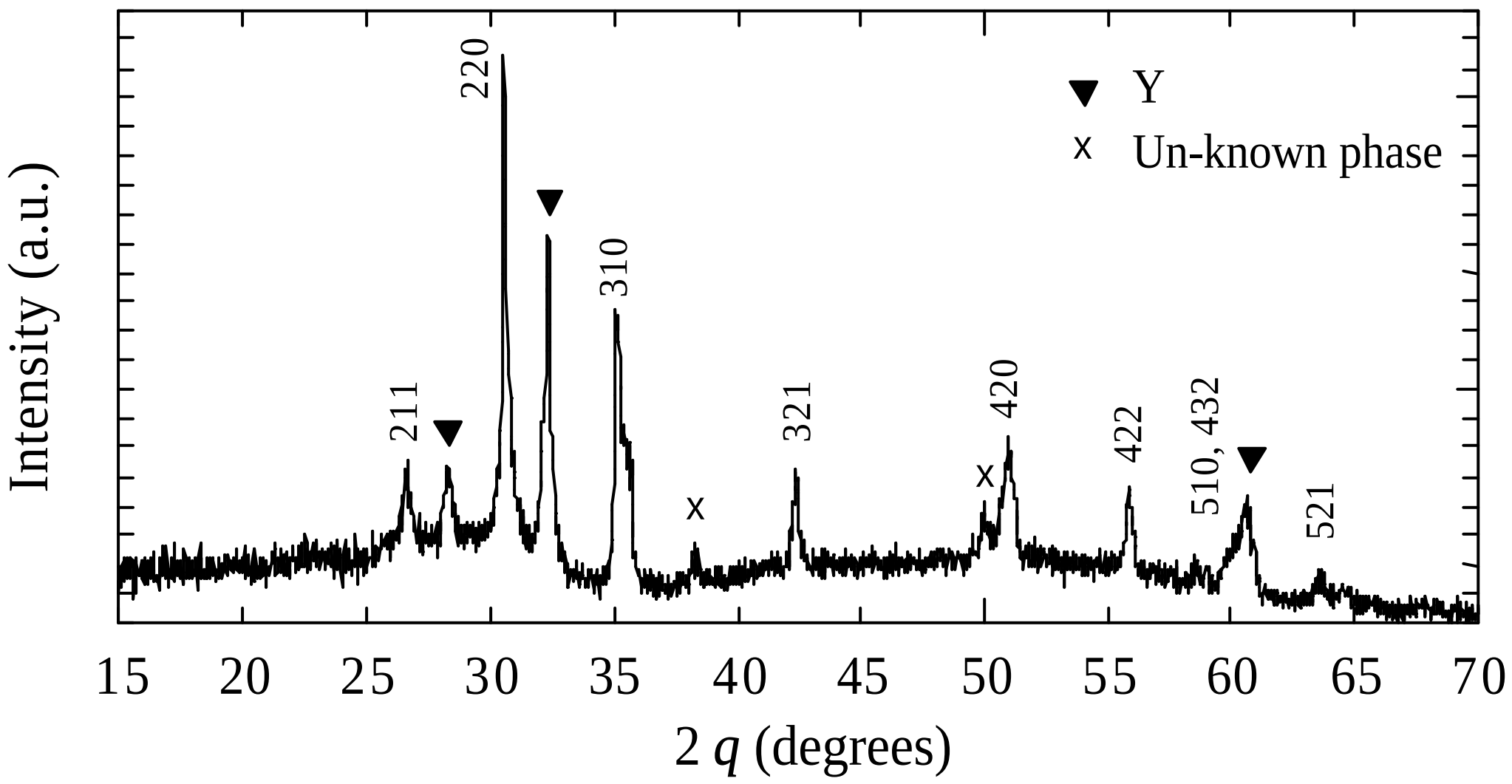

Fig. 1

Nakane *et al.*

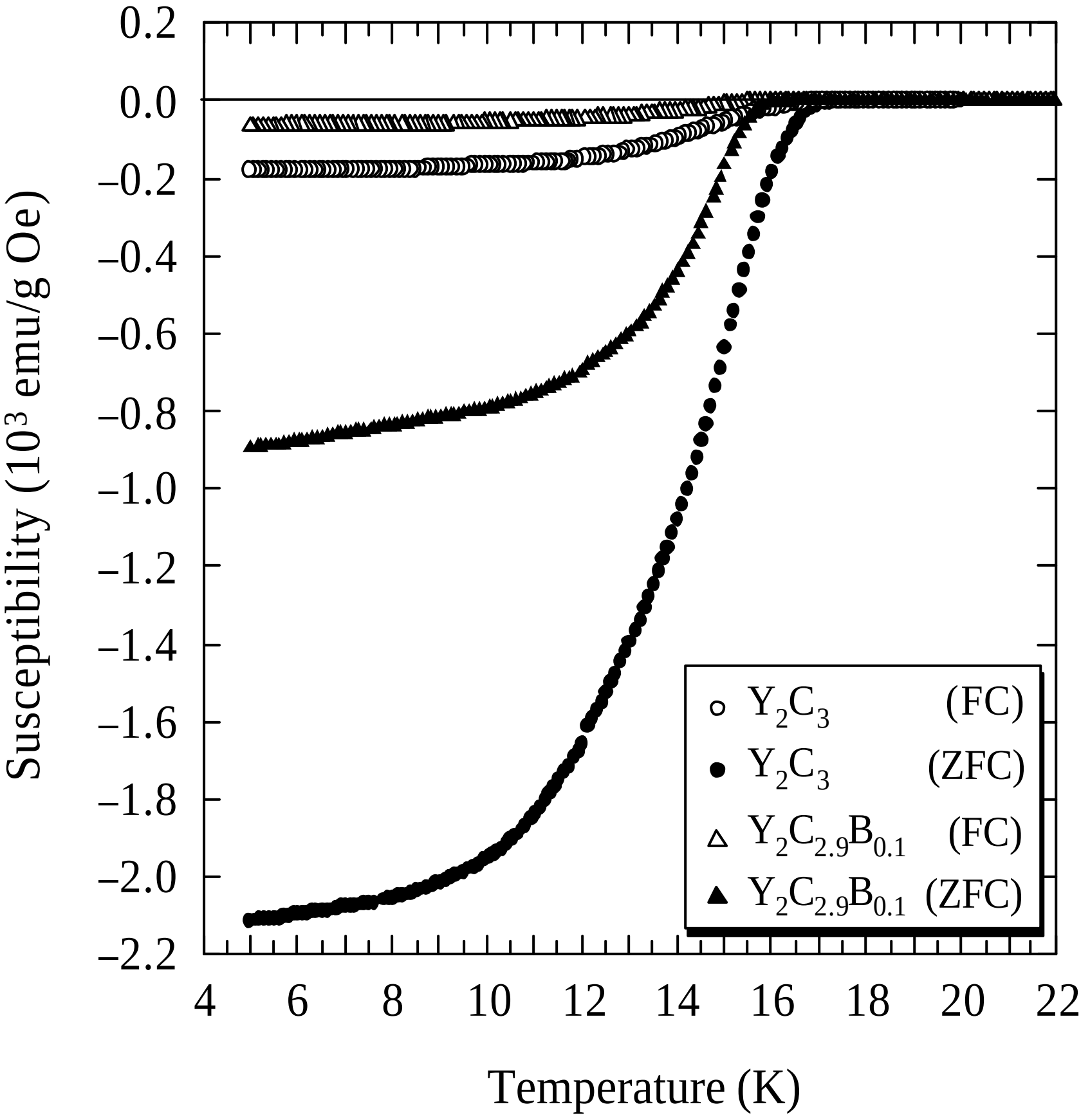

Fig. 2

Nakane *et al.*

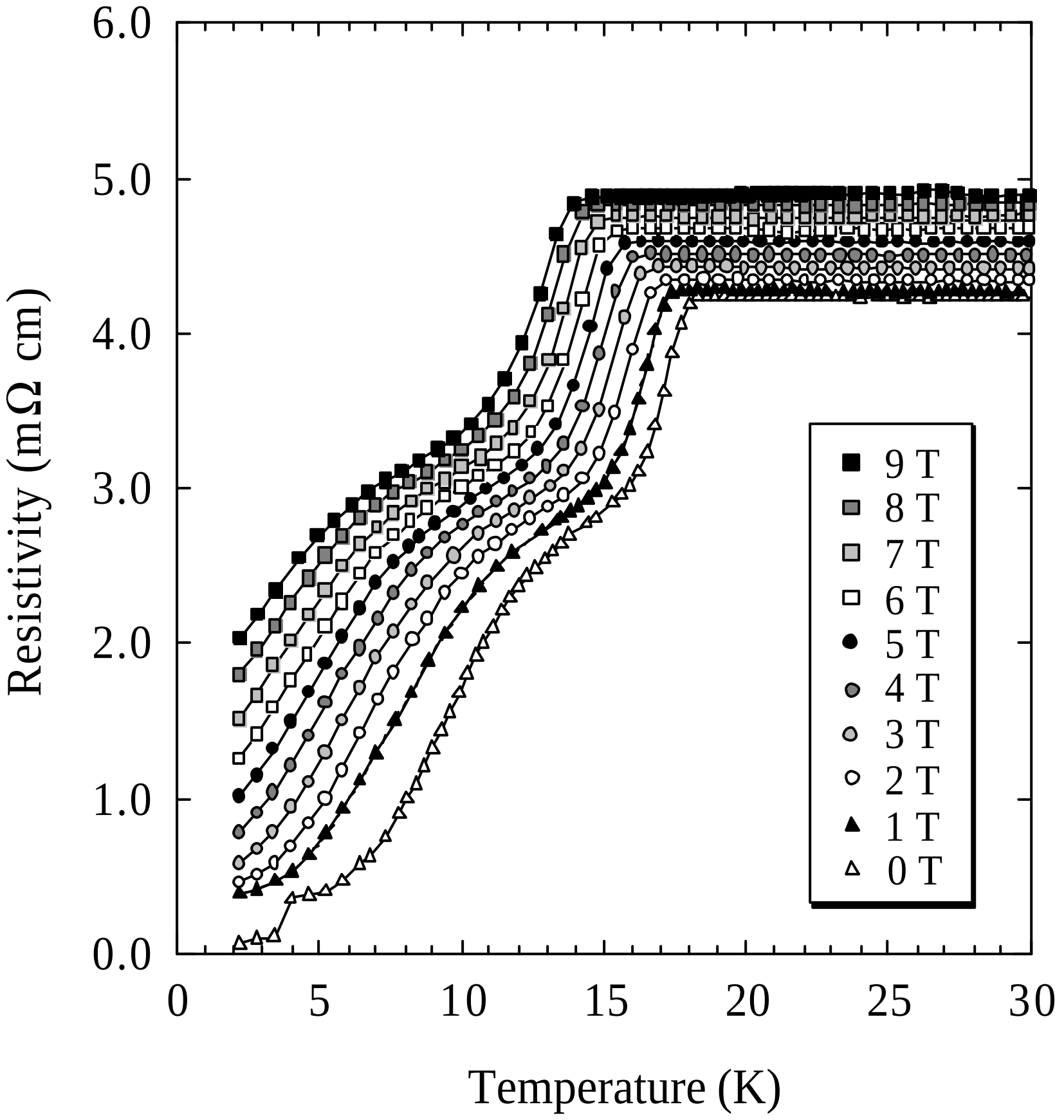

Fig. 3

Nakane *et al.*

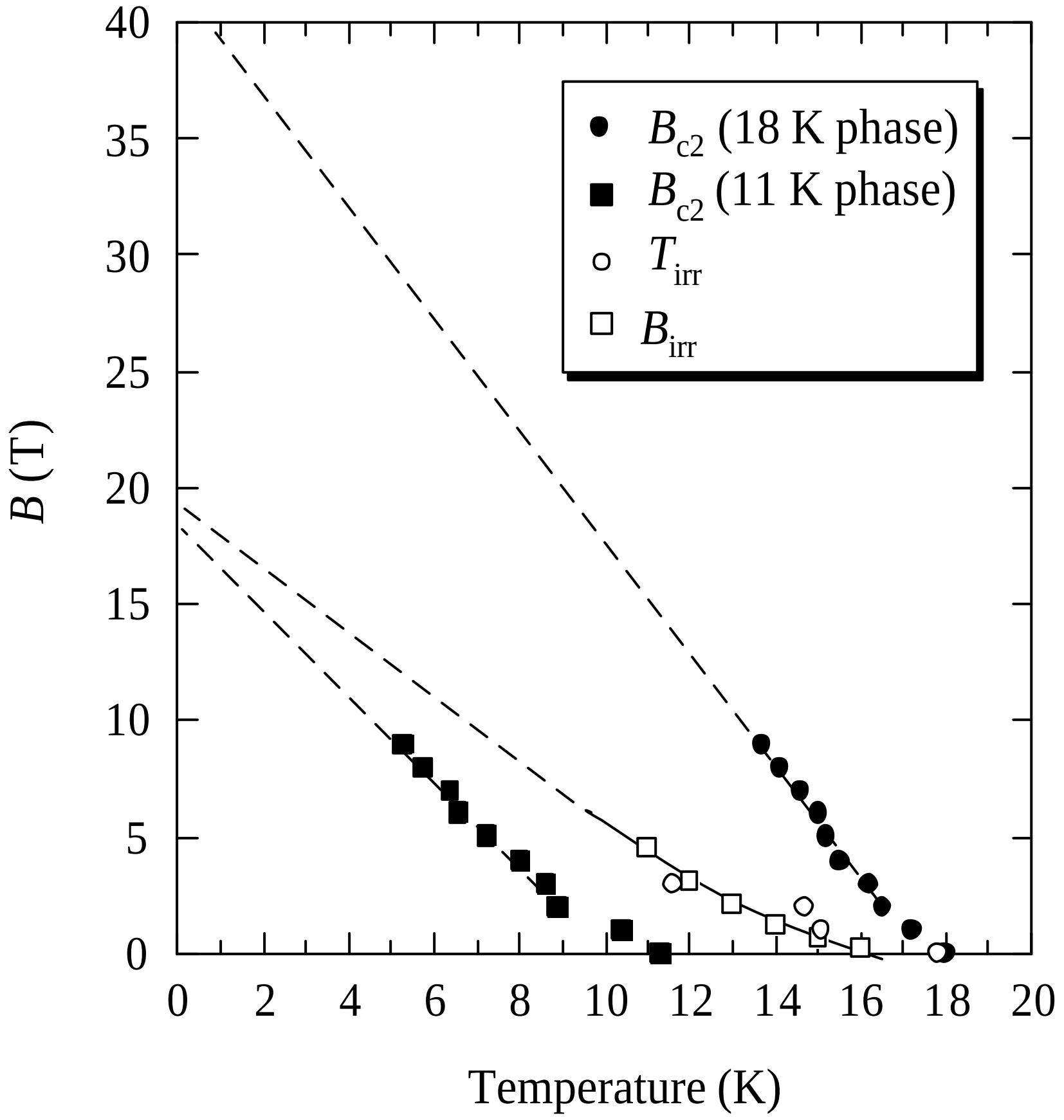

Fig. 4

Nakane *et al.*